\documentclass[11pt]{article}
\usepackage{latexsym}
\usepackage{amssymb}
\usepackage[dvips]{graphicx}
\usepackage{epsfig}
\usepackage{rotating}
\usepackage{amsfonts}
\usepackage{amsmath}

\usepackage{hyperref}
\usepackage{pstricks}

\newtheorem{definition}{Definition}[section]

\newtheorem{lemma}{Lemma}[section]
\newtheorem{theorem}{Theorem}[section]

\begin{document}

\begin{titlepage}
\begin{flushright}
ICMPA-MPA/2009/24\\
LPT-2009-93
\end{flushright}

\vspace{20pt}

\begin{center}

{\Large\bf Bosonic Colored Group Field Theory}
\vspace{20pt}

Joseph Ben Geloun$^{a,c,d,*}$, Jacques Magnen$^{b,\dag}$ and Vincent Rivasseau $^{a,\ddag}$

\vspace{15pt}

$^{a}${\sl Laboratoire de Physique Th\'eorique, Universit\'e Paris XI}\\
{\sl 91405 Orsay Cedex, France}\\
\vspace{10pt}
$^{b}${\sl Centre de Physique Th\'eorique, Ecole Polytechnique}\\
{\sl 91128 Palaiseau Cedex, France}\\
\vspace{10pt}
$^{c}${\sl International Chair in Mathematical Physics and Applications (ICMPA--UNESCO Chair),\\
University of Abomey--Calavi, 072 B. P. 50, Cotonou, Republic of Benin}\\
\vspace{10pt}
$^{d}${\sl D\'epartement de Math\'ematiques et Informatique,\\
Facult\'e des Sciences et Techniques, Universit\'e Cheikh Anta Diop, Senegal}

\vspace{20pt}

E-mail:  $^{*}${\em bengeloun@sun.ac.za},\quad $^{\dag}${\em magnen@cpht.polytechnique.fr},
\quad $^{\ddag}${\em rivass@th.u-psud.fr }

\vspace{10pt}

\begin{abstract}
\noindent
Bosonic colored group field theory is considered. Focusing first on dimension
four, namely the colored Ooguri group field model,  the main properties of Feynman graphs are studied.
This leads to a theorem on optimal perturbative bounds of Feynman amplitudes
in the ``ultraspin'' (large spin) limit. The results are generalized in any dimension.
Finally integrating out two colors we write a new representation
which could be useful for the constructive analysis of this type of models.
\end{abstract}

\end{center}

\noindent  Pacs numbers:   04.60.-m, 04.60.Pp

\end{titlepage}

\setcounter{footnote}{0}

\section{Introduction}
\label{Intro}

Group field theories (GFT's) or quantum field theories over group manifolds
were introduced in the beginning of the 90's  \cite{boul,oogur} as generalizations
of matrix and tensor field theories  \cite{Freidel,oriti}.
They are currently under active investigation as they provide one of the most complete
definition for spin foam models, themselves acknowledged as good candidates
for a background independent quantum theory of gravity \cite{rovel,thiem}.
Also they are more and more studied {\it per se}
as a full  theory of quantum gravity \cite{oriti,fgo,MNRS}.
GFT's are gauge invariant theories characterized by a non-local interaction
which pairs the field arguments in a dual way to the gluing of simplicial complexes.
Hence, the Feynman diagrams of a GFT are fat graphs,
and their duals are triangulations of (pseudo)manifolds made of
vertices, edges, faces and higher dimensional simplices
discretizing a particular spacetime.
The first and simplest GFT's  \cite{boul,oogur} have
Feynman amplitudes which are products of delta functions on the
holonomies of a group connection associated to the faces of the Feynman diagram.
These amplitudes are therefore discretizations of the topological BF theory.

But the GFT is more than simply its perturbative Feynman amplitudes.
A quantum field theory formulated by a functional integral
also assigns a precise weight to each Feynman graph,
and should resum all such graphs. In particular,
it sums over different spacetime topologies\footnote{
In this sense, GFT's are quantum field theories {\it of} the spacetime
and not solely {\it on} a spacetime \cite{oriti}.}
thus providing a so-called ``third quantization''. However a difficulty occurs, shared by
other discrete approaches to quantum gravity: since a GFT sums over arbitrary spacetime topologies,
 it is not clear how the low energy description of such a theory would
lead to a smooth and large manifold, namely our classical spacetime
(see \cite{fgo} for more details).

In order to overcome this difficulty we propose to
include the requirement of renormalizability
as a guide \cite{MNRS}.
The declared goal would be to find, using a novel scale analysis, which of the GFT models
leads to a physically relevant situation after passing through the renormalization group analysis.
Let us mention that a number of non-local noncommutative quantum field theories
have been successfully renormalized in the recent years
(see \cite{riv} and references therein). Even though the GFT's graphs are more complicated than
the ribbon graphs of noncommutative field theories, one can hope to extend to GFT's
some of the tools forged to renormalize these noncommutative quantum field theories.

Recent works have addressed the first steps of a renormalization program
for GFT's  \cite{fgo,MNRS} and  for spin foam models \cite{prs,barret} combining
topological and asymptotic large spin (also called ``ultraspin'') analysis of the amplitudes.
A first systematic analysis of the Boulatov model was started
in  \cite{fgo}. The authors have identified a specific class of
graphs called ``Type I'' for which a complete procedure of contraction is possible.
The exact power counting for these graphs has been established.
They also formulated the conjecture that these graphs dominate in the ultraspin
regime. The Boulatov model was also considered in \cite{MNRS}
as well as its Freidel-Louapre constructive regularization \cite{flouapr}.
Here, Feynman amplitudes have been studied in the large cutoff limit
and their optimal bounds have been found (at least for 
graphs without generalized tadpoles, see Appendix below). These bounds show that
the Freidel-Louapre model is perturbatively more divergent than the ordinary one.
At the constructive field theory level, Borel summability of the connected functions
in the coupling constant has been established via a convergent ``cactus'' expansion,
together with the  correct scaling of the Borel radius.

These two seminal works on the general scaling properties in GFT \cite{fgo,MNRS}
were restricted to three dimensions.
The purpose of this paper is to extend the perturbative bounds
of \cite{MNRS} to any dimension. A difficulty (which was overlooked
in \cite{MNRS}, see the Appendix of this paper) comes from the fact that
the power counting of the most general topological models is
governed by ``generalized tadpoles''.

In the mean time, a Fermionic colored GFT model possessing a $SU(D+1)$ symmetry
in dimension $D$  has been introduced by Gurau \cite{gurau}, together with
the homology theory of the corresponding colored graphs.
In contradistinction with the general Bosonic theory, the ``bubbles'' of this theory
can be easily identified and tadpoles simply do not occur.

Apart from the $SU(D+1)$ symmetry, these nice features are shared by the Bosonic version
of this colored model, which is the one considered in this paper.
We prove that a vacuum graph of such a theory is bounded
in dimension 4 by $K^n\Lambda^{9n/2}$,
where $\Lambda$ is the ``ultraspin'' cutoff and $n$ the number of
vertices of the graph. For any dimension $D$, similar bounds are also derived.
We also take the first steps towards the constructive analysis of the model.

The paper's organization is as follows. Section 2
is devoted to the definition of the colored models, the statement of some
properties of their Feynman graphs, and the
perturbative bounds which can be proved in any dimension.
Section 3 further analyses the model by integrating out two particular colors.
This leads to a Matthews-Salam formulation which reveals an
interesting hidden positivity of the model encouraging for a constructive analysis.
A conclusion is provided in Section 4 and an Appendix discusses the not yet understood
case of graphs with tadpoles in general GFT's.

\section{Perturbative bounds of colored models}
\label{Sect2}

In this section, we introduce the colored Ooguri model or colored $SU(2)$ BF theory
in four dimensions\footnote{$SU(2)$ is chosen here for simplicity.
The $SO(4)$ or $SO(3,1)$ $BF$ theories could be treated along the same lines.
Supplemented with Plebanski constraints they are a
starting point for four dimensional quantum gravity.}. 
Some useful (Feynman) graphical properties are stated
and allow us to bound a general
Feynman amplitude. These bounds
are then generalized to any dimension.

\subsection{The colored Ooguri model}
\label{subsect21}

The dynamical variables of a $D$ dimensional GFT are fields, defined over
$D$ copies of a group $G$. For the moment, let us specialize
to $D=4$ and $G=SU(2)$, hence to Ooguri-type models.

In the colored Bosonic model, these fields are themselves $D+1$
complex valued functions $\phi^\ell$, $\ell=1,2,\ldots,D +1=5$.
The upper index $\ell$ denotes the color index of the field $\phi=(\phi^1, \ldots, \phi^5)$.
The fields are required to be invariant under the ``diagonal'' action
of the group $G$
\begin{eqnarray}
&&\phi^\ell(g_1h,g_2h,g_3h,g_4h)=\phi^\ell(g_1,g_2,g_3,g_4),\;\;\;  h\in G,\cr
&&\bar\phi^\ell(g_1h,g_2h,g_3h,g_4h)=\bar\phi^\ell(g_1,g_2,g_3,g_4),\;\;\; h\in G,
\label{gi}
\end{eqnarray}
but are not symmetric  under any permutation of their arguments.
We will use the shorthand notation
$\phi^\ell_{\alpha_1,\alpha_2,\alpha_3,\alpha_4}
:=\phi^\ell(g_{\alpha_1},g_{\alpha_2},g_{\alpha_3},g_{\alpha_4})$.

The dynamics is traditionally written in terms of an action
\begin{eqnarray}
S[\phi] &:= &  \int \,  \prod_{i=1}^4dg_i \;\;\;  \sum_{\ell=1}^5\, {\bar\phi}^\ell_{1,2,3,4}\;\;
 \phi^\ell _{4,3,2,1} + \lambda_1
\int\prod_{i=1}^{10}dg_i\
\phi^1_{1,2,3,4}\; \phi^5_{4,5,6,7}\; \phi^4_{7,3,8,9}\; \phi^3_{9,6,2,10}\; \phi^2_{10,8,5,1} \cr
&&+ \lambda_2
\int\prod_{i=1}^{10}dg_i\
{\bar\phi}^1_{1,2,3,4}\; {\bar\phi}^5_{4,5,6,7}\; {\bar\phi}^4_{7,3,8,9}\; {\bar\phi}^3_{9,6,2,10}
\; {\bar\phi}^2_{10,8,5,1}
\label{action}
\end{eqnarray}
supplemented by the gauge invariance constraints (\ref{gi}),
${\bar\phi}^\ell \phi^\ell $ is a quadratic mass term.
Integrations are performed over copies of $G$ using products
of invariant Haar measures $dg_i$ of this group and $\lambda_{1,2}$ are coupling
constants.

The initial model of this type was Fermionic \cite{gurau}, hence the fields were
complex Grassmann variables $\psi^\ell $ and $\bar \psi^\ell$.
The corresponding monomials $\psi^1\psi^2\psi^3\psi^4\psi^5$ and $\bar\psi^1\bar\psi^2\bar\psi^3\bar\psi^4\bar\psi^5$
are $SU(5)$ invariant.
In the Bosonic model, this invariance is lost
as the monomials are only invariant under transformations with permanent 1,
which do not form a group.

More rigorously one should consider the gauge invariance constraints as part of the
propagator of the group field theory. The partition function of this Bosonic 
colored model is then rewritten as,
\begin{equation}
Z(\lambda_1,\lambda_2)= \int  d \mu_{C}[\bar\phi,\phi]\; e^{-\lambda_1 T_1[\phi]
 - \lambda_2 T_2[\bar\phi]},
 \label{parti}
\end{equation}
where $T_{1,2}$ stand for the interaction parts in the action (\ref{action})
associated with the $\phi$'s and $\bar\phi$'s respectively,
and  $d \mu_{C}[\bar\phi,\phi]$ denotes the degenerate Gaussian measure
(see a concise appendix in \cite{MNRS})
which, implicitly, combines the (ordinary not well defined) Lebesgue measure of fields
$D[\bar\phi,\phi]=\prod_\ell d\bar\phi^\ell d\phi^\ell$, the gauge invariance
constraint (\ref{gi}) and the mass term. Hence  $d \mu_{C}[\bar\phi,\phi]$
is associated with the covariance (or propagator) $C$ given by
\begin{equation}
\int \bar\phi_{1,2,3,4}^\ell \; \phi_{4',3',2',1'}^{\ell'}\; 
d\mu_C[\bar\phi,\phi] =  C_{\ell\ell'}(g_1,g_2,g_3,g_4;g'_4,g'_3,g'_2,g'_1)= 
\delta_{\ell\ell'}\int dh\, \prod_{i=1}^4\delta(g_i h (g'_i)^{-1}).
\end{equation}
As in the ordinary GFT situation, the covariance $C$, a bilinear form, can also be considered 
as an operator, which is a projector (satisfying $C^2=C$) and acts on the fields as
follows:
\begin{eqnarray}
[ C\phi ]^\ell_{1,2,3,4} = \sum_{\ell'}\int dg'_i \;C_{\ell\ell'}(g_1,g_2,g_3,g_4;g'_4,g'_3,g'_2,g'_1)\; \phi^{\ell'}_{4',3',2',1'}
=\int  dh \; \phi^\ell (g_4h,g_3h,g_2h,g_1h).
\end{eqnarray}

\begin{figure}
\begin{center}
      \centering
\includegraphics[angle=0, width=4.75cm, height=2.2cm]{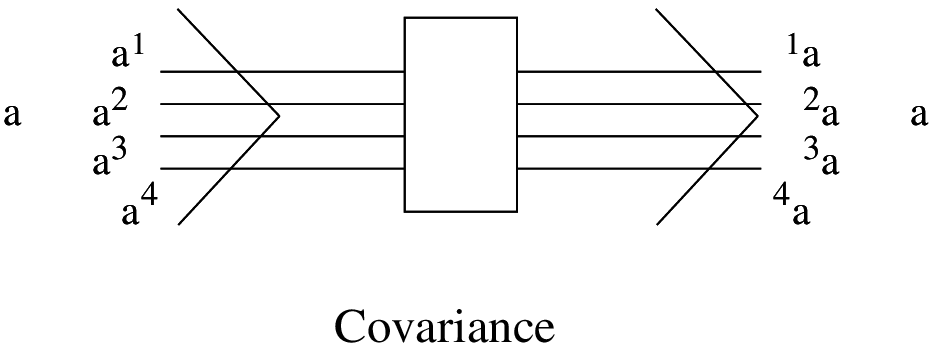}
\caption{
{\small The propagator or covariance of the colored model. }}
\centering
  \centering
\includegraphics[angle=0, width=14cm , height=6 cm]{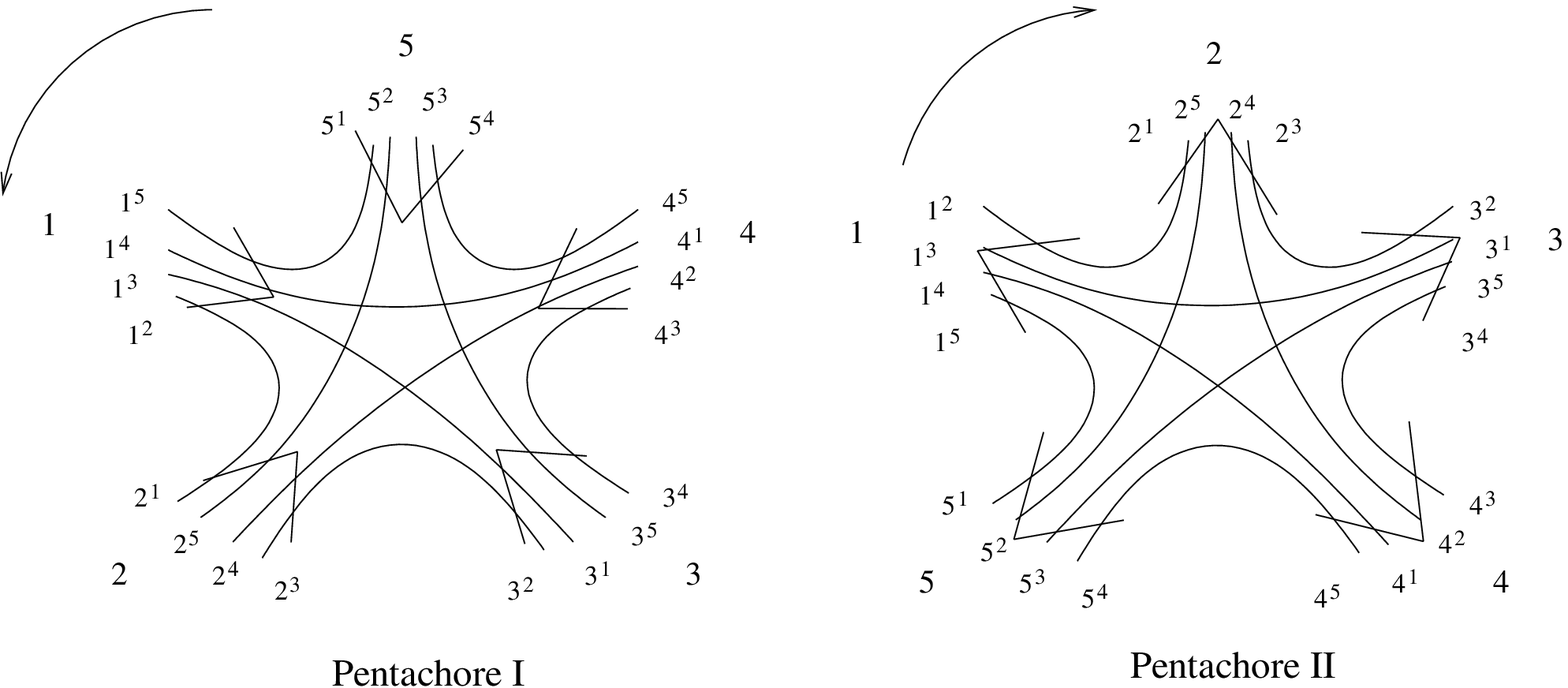}
\caption{{\small The vertices of the colored model: Pentachores I and II are associated with
interactions of the form $\phi^5$ and $\bar\phi^5$, respectively.}}
\end{center}
\end{figure}

\subsection{Properties of Feynman diagrams}
\label{subsect22}

In a given dimension $D$, Feynman graphs of a GFT are dual to $D$ dimensional
simplicial complexes triangulating a topological spacetime.
In this subsection, we illustrate the particular features of this duality
in the above colored model.

Associating  the field $\phi^\ell$ to a tetrahedron or 3-simplex with
its group element arguments $g_i$ representing its faces (triangles),
then it is well known that the order of arguments of the fields in the
quintic interactions (\ref{action})
follows the pattern of the gluing of the five tetrahedra $\ell=1,...,5$, along one
of their faces in order to build a 4-simplex or pentachore (see Fig.2).
Besides, the propagator can be seen as
the gluing rule for two tetrahedra belonging to two neighboring pentachores.
For the BF theory this gluing is made so that each face is flat.
In the present situation, the model (\ref{action}) adds new features in the theory:
the fields are complex valued and colored.
Consequently, we can represent the complex nature of the fields by a specific orientation
of the propagators (big arrows in Fig.1) and the color feature of the fields is reflected
by a specific numbering (from 1 to 5, see Fig.2) of the legs of the vertex.
Hence,  the only admissible propagation should be
between a field $\bar\phi^\ell$ and $\phi^\ell$ of the same color index, and ``physically''
it means that only tetrahedra of the same color belonging to two neighboring pentachores
can be glued together. Finally, given a propagator with color $\ell=1,2,...,5$, it has itself sub-colored lines
(called sometimes ``strands'') that we write cyclically $\ell^{j}$, and the gluing also respects these
subcolors (because our model does not include any permutation symmetry
of the strands).

Let us enumerate some useful properties of the colored graphs.
The following lemmas hold.
\begin{lemma}
\label{lem1}
Given a $N$-point graph with  $n$ internal vertices,
if one color is missing on the external legs, then
\begin{enumerate}
\item [(i)] $n$ is a even number;
\item[(ii)] $N$ is also even and external legs have colors which appear in pair.
\end{enumerate}
 \end{lemma}
{\bf Proof.} Let us consider a $N$-point graph with $n$ internal vertices.
Consequently, one has $n$ fields for each color.
Since one color is missing on the external legs,
and since by parity, any contraction creating an internal line consumes
two fields of the same color, the full contraction process for that missing color
consumes all the $n$ fields of that color and an even number of fields.
Hence $n$ must be even. This proves the point $(i)$.

We prove now the point $(ii)$. We know that the number $n$ of internal vertices is even.
Now, if a color on the external legs appears an odd number of times,
the complete internal contraction for that color would involve an odd number
of fields hence would be impossible. $\square$

An interesting corollary is that for $N < D+1$ the conclusions of Lemma \ref{lem1}
must hold. In particular the colored theory in dimension $D$ has no odd point functions
with less than $D$ arguments. For example, the colored Ooguri model in four dimensions has no one and three 
point functions. This property is reminiscent of ordinary even field theories like
the $\phi_4^4$ model which has also neither one nor three point functions. It may simplify
considerably the future analysis of renormalizable models of this type.

We recall that in a colored GFT model,
\begin{enumerate}
\item[(i)] a face (or closed cycle) is bi-colored with an even number of lines;
\item[(ii)] a chain (opened cycle) of length $>1$ is bi-colored.
\end{enumerate}
This is in fact the definition of a face in \cite{gurau} but can also be easily deduced
from the fact that each strand in a line $l_a^{(0)}$
joining a vertex $V^{(0)}$ to a vertex $V^{(1)}$
possesses a double label that we denote by $a^{b}$ (see Fig. 2):
$a$ is the color index of  $l_a^{(0)}$ and $b$ denotes
the color of the line $l_b^{(1)}$ after the vertex $V_1$,
where the strand $a^b$ will propagate. In short, a vertex connects
in a unique way,
the strand $a^b$ of the line $l_a$ to the strand $b^a$
of the line $l_b$. The point is that, in return, the strand $b^a$
belonging to $l_b^{(1)}$ can only be connected to a strand
of the form $a^b$ in a line $l_a^{(2)}$ through another vertex $V^{(2)}$.
Hence only chains of the bi-colored form
$a^b b^a a^b b^a\ldots b^a$ (for closed cycles) and
$a^b b^a a^b b^a \ldots a^b$ (for open chains) could be obtained.
Closed cycles include clearly an even number of lines.   $\square$

\begin{definition}
A generalized tadpole is a $(N<5)$-point graph with only one external vertex (see Fig.3).
\end{definition}
In the case of a generalized tadpole with one external leg, that external leg must contract to a 
single vertex, creating a new tadpole but with four external legs.
\begin{figure}
 \centering
       \begin{minipage}[t]{0.6\textwidth}
     \centering
\includegraphics[angle=0, width=8cm, height=3cm]{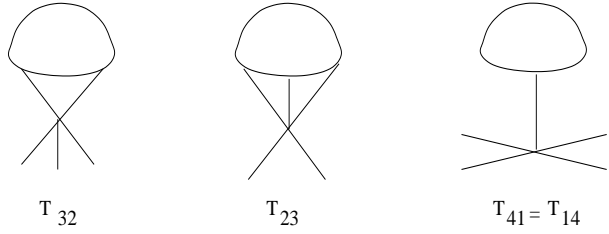}
\caption{
{\small Generalized tadpoles. }}
\end{minipage}
\end{figure}
\begin{theorem}
\label{theo1}
 There is no generalized tadpole in the colored Ooguri group field model.
\end{theorem}
{\bf Proof.} By definition, a generalized tadpole is a $(N<5)$-point graph,
meaning that at least one of the colors is missing on the external legs.
Then, Lemma \ref{lem1} tells us that $N$ should be even ($N=2$ or $4$)
and the external colors appear in pairs. Having in mind that, still by definition,
an ordinary vertex of the colored  theory has no common color
in its fields, the external vertex of a generalized tadpole
having colors appearing in pairs cannot be an ordinary colored vertex.
$\square$

Hence, the colored Ooguri model has no generalized tadpole and so, {\it a fortiori},
no tadpole. This property does not actually depend
on the dimension.

The same strategy as in \cite{MNRS} is used
in order  to determine the types of vertex operators from which the
Feynman amplitude of a general vacuum graph will be bounded in the next subsection. 
We first start by some definitions.

\begin{definition}
\begin{enumerate}
\item[(i)] A set $A$ of vertices of a graph $\mathcal{G}$ is called connected if the subgraph made of these vertices 
and all their inner lines (that is all the lines starting and ending at a vertex of $A$) is connected. 
\item[(ii)]
An $(A,B)$-cut of a two-point connected graph $\mathcal{G}$ with two external vertices $v_A$ and $v_B$
is a partition of the vertices of $\mathcal{G}$ into two subsets $A$ and $B$
such that $v_A \in A$, $v_B \in B$ and $A$ and $B$ are still connected.
\item[(iii)] A line joining a vertex of $A$ to a vertex of $B$ in the graph is called a frontier line
for the $(A,B)$-cut.
\item[(iv)] A vertex of $B$ is called a frontier vertex with respect to the cut if there is a frontier
line attached to that vertex.
\item[(v)] An exhausting sequence of cuts for a connected  two-point graph $\mathcal{G}$ of order $n$
is a sequence $A_0 =\emptyset  \varsubsetneqq
A_1  \varsubsetneqq A_2 \varsubsetneqq \cdots  \varsubsetneqq A_{n-1}  \varsubsetneqq A_n =\mathcal{G}$
such that $(A_p, B_p := \mathcal{G}\setminus A_p)$ is a cut of $\mathcal{G}$ for any $p=1,  \cdots , n-1$.
\end{enumerate}
\end{definition}
Given a graph $\mathcal{G}$, an exhausting sequence of cuts is a kind of total ordering
of the vertices of $\mathcal{G}$, such that each vertex can be `pulled' successively
through a `frontier' from one part $B$ to another part $A$,
and this without disconnecting $A$ or $B$.
\begin{lemma}
\label{lem3}
Let $\mathcal{G}$ be a colored connected two-point graph. There exists an exhausting sequence of
cuts for $\mathcal{G}$.
\end{lemma}
The proof can be worked out by induction along the lines of a similar lemma in \cite{MNRS}.
The main ingredient for this proof in the colored model is the absence of generalized tadpoles
as established by Theorem \ref{theo1}. We reproduce here this proof in detail in the four
dimensional case for completeness, but the result holds in any dimension.

\noindent{\bf Proof  of Lemma \ref{lem3}.}
Let us consider a two-point graph $\mathcal{G}$ with $n$ vertices and assume that
a sequence $A_0 =\emptyset  \varsubsetneqq
A_1  \varsubsetneqq A_2 \varsubsetneqq \cdots A_p$ and its corresponding sequence
$\{B_{j=0,1,\ldots,p}\}$, have
been defined for $0\le p<n-1$, such that for all $j=0,1,\ldots, p$, $(A_j,B_j)$ is a cut for $\mathcal{G}$.
Then another frontier vertex $v_{p+1}$ has to be determined such that $A_{p+1}=A_{p}\cup\{v_{p+1}\}$
and $B_{p+1}=\mathcal{G} \setminus A_{p+1}$ define again a cut for $\mathcal{G}$.

 Let us consider a tree $T_p$ with a fixed root $v_B$
which spans the remainder set of vertices $B_p$ and give them a partial ordering.
The set $B_p$ being finite,  we can single out a maximal frontier vertex $v_{max} $
with respect to that ordering, namely a frontier vertex such that there is no other
frontier vertex in the ``branch above $v_{max}$" in $T_p$.

We prove first that $v_{max} \ne v_B$. The proposition $v_{max} = v_B$ would imply that
$v_B$ is the only frontier vertex left in $B_p$. The vertex $v_B$ has four internal lines
and therefore two cases may occur:
(1) all these are frontier lines then this means that $\{v_B\}=B_p$ which contradicts
the fact that $p<n-1$; (2) not all lines are frontier lines which implies that
some of these lines span a generalized tadpole.
This possibility does not occur by Theorem \ref{theo1}.

Let us then assume that $v_{max} \ne v_B$ and choose $v_{p+1}=v_{max}$.
We would want to prove that the set  $B_{p+1} = B_p \setminus\{v_{p+1}\}$
is still connected through internal lines.
Cutting $\ell_{p+1}$, the unique link  between  $v_{p+1}$ and $v_B$ in $T_p$
splits the tree into two connected components. We call $R_p$ the part containing $v_B$,
and $S_p$ the other  one. Note that $S_p$ becomes a rooted tree with root and only frontier vertex $v_{p+1}$
(remember that $v_{p+1}$ is maximal). From the property of $v_{p+1}$ to be a frontier vertex,
one deduces that it has at most four lines in $B_p$ (and so at least one frontier line) and
hence there are at most three lines from $v_{p+1}$ to $B_{p+1}$ distinct from $\ell_{p+1}$.
Since the tree $R_p$ does not have any line hooked to $v_{p+1}$, its lines remain inner
lines of $B_{p+1}$ confining all of its vertices to a single connected component of $B_{p+1}$.
Assume that $B_{p+1}$ is not connected, this would imply first, that $S_p$ contains other vertices
than $v_{p+1}$. Second, removing the root $v_{p+1}$ from $S_p$ and the at most three lines hooked to it,
one, two, or three connected components would be obtained. These component are made of the vertices
of $S_p\setminus \{v_{p+1}\}$ plus their inner lines, which no longer hook to $R_p$
through inner lines of $B_{p+1}$. Due to the fact that these components have no frontier vertices
hence no other frontier lines, it would mean that they must have been
hooked to the total graph $\mathcal{G}$ through at most three lines  from $v_{p+1}$ to $B_{p+1}$ distinct from
$\ell_{p+1}$, hence they would have been a generalized tadpole which is not admissible. $\square$

\subsection{Perturbative bounds}
\label{subsect23}

\begin{figure}
\centering
    \begin{minipage}[t]{0.6\textwidth}
\centering
\includegraphics[angle=0, width=8cm, height=3.5cm]{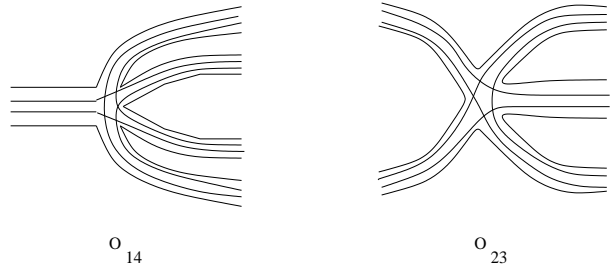}
\caption{
{\small The vertex operators. }}
\end{minipage}
\end{figure}

To begin with the study of perturbative bounds of the Feynman graphs,
let us introduce a cutoff of the theory in order to define a regularized version
of the partition function $Z(\lambda_1,\lambda_2)$ (\ref{parti}).
We then truncate the Peter-Weyl field expansion as
\begin{equation}
\phi^\ell_{1,2,3,4} =
\sum_{{j}_1,j_2,j_3,j_4}^{\Lambda} \text{tr}
\left(\Phi_{j_1,j_2,j_3,j_4} D^{j_1}(g_1) D^{j_2}(g_2) D^{j_3}(g_3) D^{j_4}(g_4)\right),
\end{equation}
where the summation is over the spin indices $j_{1,...,4}$, up to $\Lambda$,
$D^{j}(g)$ denotes the $(2j+1)$-dimensional matrix representing $g$
and $\Phi_{j_1,j_2,j_3,j_4}$ are the corresponding modes.
On the group, the delta function with cutoff is of the form
$\delta_\Lambda(h) = \sum_{j}^\Lambda (2j+1)\, \text{tr}\, D^j (h)$.
This function behaves as usual as
$\int dg \: \delta_\Lambda (h g^{-1})\delta_\Lambda (gk^{-1})=\delta_\Lambda (hk^{-1})$,
$\int dg \delta(gh)=1$, and diverges as $\sum_j^\Lambda j^2 \sim \Lambda^3$.

\begin{figure}[t]
\begin{center}
\scalebox{0.7}
{
\begin{pspicture}(0,-2.515)(11.915,2.515)

\psline[linewidth=0.03cm](3.3,-1.7)(3.5,-1.7)
\psbezier[linewidth=0.03](3.5,-1.7)(3.7,-2.3)(4.1,-2.3)(4.3,-1.7)
\psbezier[linewidth=0.03](3.5,-1.7)(3.7,-1.1)(4.1,-1.1)(4.3,-1.7)

\psbezier[linewidth=0.03](3.5,-1.7)(3.7,-2)(4.1,-2)(4.3,-1.7)
\psbezier[linewidth=0.03](3.5,-1.7)(3.7,-1.5)(4.1,-1.5)(4.3,-1.7)

\psline[linewidth=0.03cm](4.3,-1.7)(4.5,-1.7)

\psbezier[linewidth=0.03](4.5,-1.7)(4.7,-2.3)(5.1,-2.3)(5.3,-1.7)
\psbezier[linewidth=0.03](4.5,-1.7)(4.7,-1.1)(5.1,-1.1)(5.3,-1.7)

\psbezier[linewidth=0.03](4.5,-1.7)(4.7,-2)(5.1,-2)(5.3,-1.7)
\psbezier[linewidth=0.03](4.5,-1.7)(4.7,-1.5)(5.1,-1.5)(5.3,-1.7)

\psline[linewidth=0.03cm](5.3,-1.7)(5.5,-1.7)

\psbezier[linewidth=0.03](5.5,-1.7)(5.7,-2.3)(6.1,-2.3)(6.3,-1.7)
\psbezier[linewidth=0.03](5.5,-1.7)(5.7,-1.1)(6.1,-1.1)(6.3,-1.7)

\psbezier[linewidth=0.03](5.5,-1.7)(5.7,-2)(6.1,-2)(6.3,-1.7)
\psbezier[linewidth=0.03](5.5,-1.7)(5.7,-1.5)(6.1,-1.5)(6.3,-1.7)

\psline[linewidth=0.03cm](6.3,-1.7)(6.5,-1.7)

\psdots[dotsize=0.08](6.7,-1.7)
\psdots[dotsize=0.08](6.9,-1.7)
\psdots[dotsize=0.08](7.1,-1.7)
\psdots[dotsize=0.08](6.7,-1.7)
\psdots[dotsize=0.08](6.9,-1.7)
\psdots[dotsize=0.08](7.3,-1.7)
\psdots[dotsize=0.08](7.5,-1.7)
\psdots[dotsize=0.08](7.7,-1.7)
\psdots[dotsize=0.08](7.9,-1.7)
\psdots[dotsize=0.08](8.1,-1.7)

\psline[linewidth=0.03cm](8.3,-1.7)(8.5,-1.7)

\psbezier[linewidth=0.03](8.5,-1.7)(8.7,-2.3)(9.1,-2.3)(9.3,-1.7)
\psbezier[linewidth=0.03](8.5,-1.7)(8.7,-1.5)(9.1,-1.5)(9.3,-1.7)

\psbezier[linewidth=0.03](8.5,-1.7)(8.7,-2)(9.1,-2)(9.3,-1.7)
\psbezier[linewidth=0.03](8.5,-1.7)(8.7,-1.1)(9.1,-1.1)(9.3,-1.7)

\psline[linewidth=0.03cm](9.3,-1.7)(9.5,-1.7)

\psbezier[linewidth=0.03](9.5,-1.7)(9.7,-2.3)(10.1,-2.3)(10.3,-1.7)
\psbezier[linewidth=0.03](9.5,-1.7)(9.7,-1.5)(10.1,-1.5)(10.3,-1.7)

\psbezier[linewidth=0.03](9.5,-1.7)(9.7,-2)(10.1,-2)(10.3,-1.7)
\psbezier[linewidth=0.03](9.5,-1.7)(9.7,-1.1)(10.1,-1.1)(10.3,-1.7)

\psline[linewidth=0.03cm](10.3,-1.7)(10.5,-1.7)

\psbezier[linewidth=0.03](3.3,-1.7)(1.9,-1.7)(3.5,-2.5)(6.9,-2.5)
\psbezier[linewidth=0.03](10.5,-1.7)(11.9,-1.7)(10.3,-2.5)(6.9,-2.5)

\psdots[dotsize=0.08](3.3,1.1)
\psdots[dotsize=0.08](4.5,1.1)
\psbezier[linewidth=0.03](3.3,1.1)(3.7,0.7)(4.1,0.7)(4.5,1.1)
\psline[linewidth=0.03cm](3.3,1.1)(4.5,1.1)
\psbezier[linewidth=0.03](3.3,1.1)(3.7,1.5)(4.1,1.5)(4.5,1.1)
\psbezier[linewidth=0.03](8.1,1.1)(8.5,0.7)(8.9,0.7)(9.3,1.1)
\psline[linewidth=0.03cm](5.7,1.1)(6.9,1.1)
\psbezier[linewidth=0.03](8.1,1.1)(8.5,1.5)(8.9,1.5)(9.3,1.1)
\psline[linewidth=0.03cm](9.3,1.1)(10.4,1.1)
\psdots[dotsize=0.08](7.8,1.1)
\psdots[dotsize=0.08](7.6,1.1)
\psdots[dotsize=0.08](7.4,1.1)
\psdots[dotsize=0.08](7.2,1.1)
\psbezier[linewidth=0.03](5.7,1.1)(6.1,0.7)(6.5,0.7)(6.9,1.1)
\psbezier[linewidth=0.03](5.7,1.1)(6.1,1.5)(6.5,1.5)(6.9,1.1)
\psdots[dotsize=0.08](4.5,1.1)
\psbezier[linewidth=0.03](4.5,1.1)(4.9,0.7)(5.3,0.7)(5.7,1.1)
\psbezier[linewidth=0.03](4.5,1.1)(4.9,1.5)(5.3,1.5)(5.7,1.1)
\psbezier[linewidth=0.03](9.3,1.1)(9.7,0.7)(10.1,0.7)(10.5,1.1)
\psbezier[linewidth=0.03](9.3,1.1)(9.7,1.5)(10.1,1.5)(10.5,1.1)
\psdots[dotsize=0.08](5.7,1.1)
\psdots[dotsize=0.08](6.9,1.1)
\psdots[dotsize=0.08](8.1,1.1)
\psdots[dotsize=0.08](9.3,1.1)
\psdots[dotsize=0.08](10.5,1.1)
\psbezier[linewidth=0.03](3.3,1.1)(2.3,2.1)(5.3,2.5)(6.9,2.5)
\psbezier[linewidth=0.03](10.5,1.1)(11.5,2.1)(8.5,2.5)(6.9,2.5)
\psbezier[linewidth=0.03](3.3,1.1)(2.3,0.1)(5.3,-0.3)(6.9,-0.3)
\psbezier[linewidth=0.03](10.5,1.1)(11.5,0.1)(8.5,-0.3)(6.9,-0.3)
\rput(1,1.1){ $G^{23}_n\quad=$}
\rput(1,-1.8){ $G^{14}_n\quad=$}
\end{pspicture}
}
\caption{The chains of graphs $G^{23}_{n}$ and $G^{14}_{n}$ with $2n$ vertices.}
\vspace{0.3cm}
\end{center}
\begin{center}
\centering
\includegraphics[angle=0, width=12cm, height=3.5cm]{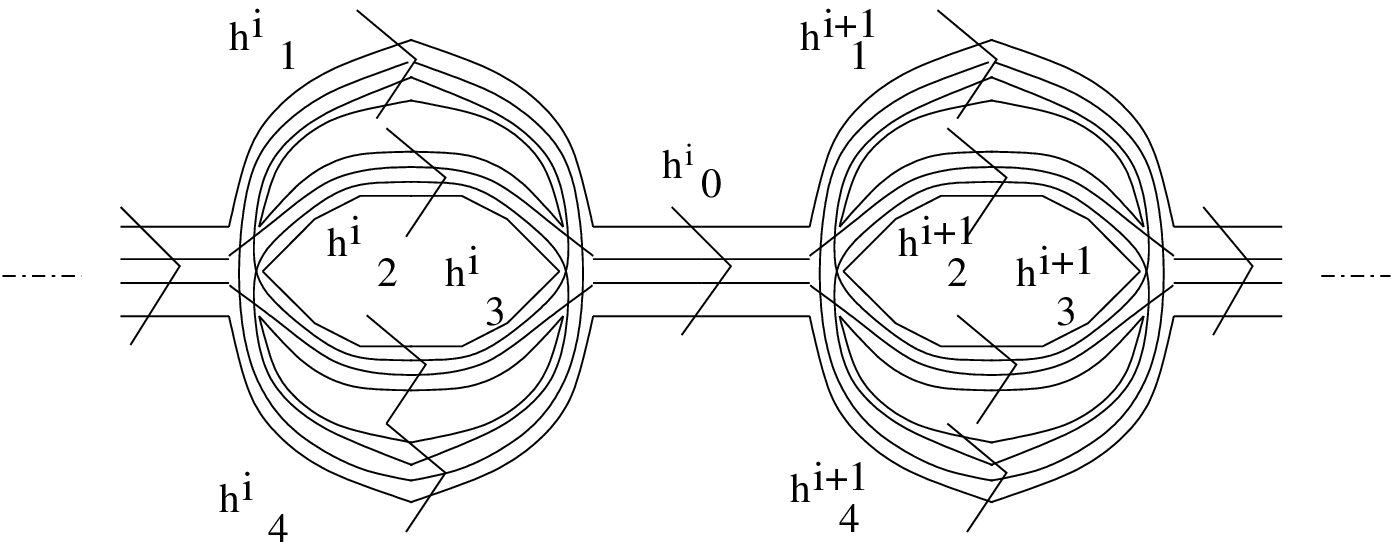}
\caption{
{\small Elements in the chain $G^{14}_n$. }}
\end{center}
\end{figure}
We mention that the bounds obtained hereafter are only valid for vacuum graphs.
For general graphs with external legs, a bound can be easily obtained by
Cauchy-Schwarz inequalities by taking the product of the $L^2$-norms of these external
legs (seen as test functions) times the amplitude of a vacuum graph.

We now prove the following theorem
\begin{theorem}
There exists a constant $K$ such that for any connected colored vacuum graph $\mathcal{G}$ of the
Ooguri model with $n$ internal vertices, we have
\begin{equation}
 |\mathcal{A}_\mathcal{G}|\leq K^n \Lambda^{9n/2 + 9}.
\end{equation}
This bound is optimal in the sense that there exists a graph $\mathcal{G}_n$ with $n$ internal vertices
 such that  $|\mathcal{A}_\mathcal{G}| \simeq  K^n \Lambda^{9n/2 + 9}$.
\end{theorem}
{\bf Proof.} Lemma \ref{lem3} shows that the vertices can be ``pulled out'' one by one
from the connected graph $\mathcal{G}$. From this procedure, we build two kinds of vertex operators
which compose the graph $\mathcal{G}$.  These are the operators $O_{14}$ and $O_{23}$ (see Fig.4) and their
adjoint acting as
\begin{eqnarray}
 O_{14}: \mathcal{H}^\Lambda_0 & \longrightarrow & \mathcal{H}^\Lambda_0\otimes
\mathcal{H}^\Lambda_0 \otimes \mathcal{H}^\Lambda_0 \otimes  \mathcal{H}^\Lambda_0 \cr
 O_{23}: \mathcal{H}^\Lambda_0  \otimes  \mathcal{H}^\Lambda_0
 & \longrightarrow & \mathcal{H}^\Lambda_0\otimes
\mathcal{H}^\Lambda_0 \otimes \mathcal{H}^\Lambda_0
\end{eqnarray}
where $\mathcal{H}^\Lambda_0:=  \mathcal{H}^\Lambda \cap \text{Im}\, C$ is a subspace of
$SU(2)$-right invariant  function belonging also to
$\mathcal{H}^\Lambda \subset L^2(SU(2)^4)$ the subspace of $L^2$ integrable
functions with $\Lambda$-truncated  Peter-Weyl expansion.

The norm of these operators can be computed according to the formula
\begin{equation}
 ||H|| = \lim_{n \to \infty} \left( \text{tr} [H^\dag H]^n\right)^{1/2n},
\end{equation}
where $H^\dag$ denotes the adjoint operator associated with $H$.
We start by the calculations of $ \text{tr}(O_{14} O_{41})^n$ using
the formula \cite{fgo}
\begin{equation}
  \text{tr}(O_{14} O_{41})^n =
\int
\prod_{ l\in \mathcal{L}_{G^{14}_n} } dh_l \prod_{  f\in \mathcal{F}_{G^{14}_n}}
\delta_\Lambda\left({\vec{\prod}_{l \in \partial f} h_l }\right)
\end{equation}
with  $G^{14}_n$ (see Fig.5 and Fig.6) the vacuum graph obtained from $ \text{tr}(O_{14} O_{41})^n$,
$\mathcal{L}_{G^{14}_n}$ its set of lines, $\mathcal{F}_{G^{14}_n}$ its
set of faces (closed cycles of strands). The oriented product in the argument of the
delta function is to be performed on the $h_l$ belonging to each oriented line
of each oriented face.  The rule is that if the orientation of the line
and the face coincide, one takes $h_l$ in the product; if the orientations disagree
one takes the value $h^{-1}_l$.
We get after a  reduction (we omit henceforth the subscript $\Lambda$ in the notation of
the truncated  delta functions)
\begin{eqnarray}
  &&\text{tr} (O_{14} O_{41})^n =\crcr
&&   \int \prod_{i=1}^n \prod_{j=0}^4 dh^i_j \left\{\;
 \delta (\prod_{i=1}^n  h^i_1h^i_0 )\delta(h^n_0 h^1_1)
 \delta (\prod_{i=1}^n  h^i_2h^i_0)\delta(h^n_0 h^1_2)
 \delta (\prod_{i=1}^n  h^i_3h^i_0)\delta(h^n_0 h^1_3)
 \delta (\prod_{i=1}^n  h^i_4 h^i_0)\delta(h^n_0 h^1_4)\right. \cr
&&\left.
\prod_{i=1}^{n} \delta(h_1^i (h_2^i)^{-1}) \delta(h_1^i (h_3^i)^{-1})
\delta(h_1^i (h_4^i)^{-1})\delta(h_2^i (h_3^i)^{-1})\delta(h_2^i (h_4^i)^{-1})
\delta(h_3^i (h_4^i)^{-1}) \right\}.
\end{eqnarray}
After integration,  it is simple to deduce that $\text{tr} (O_{14} O_{41})^n
\leq \Lambda^{9n+9}$ and the
operator norm is bounded by
\begin{eqnarray}
|| O_{14}  || \leq \Lambda^{9/2}.
\label{o14}
\end{eqnarray}
A similar calculation allows us to write $\text{tr} (O_{23} O_{32})^{2n}
\leq \Lambda^{6n+15}$  (see graph $G^{23}_n$ in Fig.5) such  that
we get the bound
\begin{equation}
|| O_{23} || \leq \Lambda^{3/2}. \qquad \qquad \square
\label{o23}
\end{equation}

The meaning of the norms of the operators $O_{14}$ and $O_{23}$ is the following:
each vertex in a graph $\mathcal{G}_n$ diverges at most as $ \Lambda^{9/2}$
which is the bound of $||O_{14}||$. Roughly speaking the amplitude of a colored graph is bounded 
by $K^{n}\Lambda^{9n/2}$ where $n$ is its number of vertices.

\subsection{D dimensional colored GFT}
\label{subsect24}

We treat in this subsection, in a streamlined analysis,
how to extend the above perturbative bounds to any dimension
$D$.

\noindent{\bf The D dimensional  GFT model.}
In dimension $D$, the action (\ref{action}) finds the following
extension
\begin{eqnarray}
 &&S^{D}[\phi] :=   \int \,  \prod_{i=1}^Ddg_i \;\;\;  \sum_{\ell=1}^{D+1}\, {\bar\phi}^\ell_{1,2,\ldots,D}\;\;
 \phi^\ell _{D,\ldots,2,1} \crcr
&&+ \lambda_1
\int\prod dg_{i^j}\;
\phi^1_{1^2,1^3,\ldots, 1^{(D+1)}}\; \phi^{(D+1)}_{(D+1)^1,(D+1)^2,(D+1)^3,\ldots,(D+1)^{D}}\;
 \phi^{D}_{D^{D+1},D^{1},D^2,\ldots, D^{D-1}}\ldots\cr
&&\qquad \ldots  \ldots \phi^3_{3^{4},3^{5},\ldots,3^{D+1}, 3^{1},3^2}\;
 \phi^2_{2^{3},2^4,\ldots,2^{D+1}, 2^{1}} \prod_{j\ne i}^{D+1} \delta(g_{i^j}(g_{j^i})^{-1})\cr
\crcr
&&+ \lambda_2
\int\prod dg_{i^j}\;
\bar\phi^1_{1^2,1^3,\ldots, 1^{(D+1)}}\; \bar\phi^{D+1}_{(D+1)^1,(D+1)^2,(D+1)^3,\ldots,(D+1)^{D}}\;
\bar \phi^{D}_{D^{D+1},D^{1},D^2,\ldots, D^{D-1}}\ldots\cr
&&\qquad \ldots  \ldots \bar\phi^3_{3^{4},3^{5},\ldots,3^{D+1}, 3^{1},3^2}\;
 \bar\phi^2_{2^{3},2^4,\ldots,2^{D+1}, 2^{1}} \prod_{j\ne i}^{D+1} \delta(g_{i^j}(g_{j^i})^{-1})\cr
\end{eqnarray}
where the complex colored fields $\phi^\ell:G^{D}\to \mathbb{C}$  will be denoted by
$\phi^\ell(g_{\ell^i},g_{\ell^j},\ldots, g_{\ell^k})=\phi^\ell_{\ell^i,\ell^j,\ldots, \ell^k}$,
and $g_{\ell^k}$ is a group element materializing the link in the vertex of two
colors $\ell$ and $k$ (the general propagator and vertex $\phi^{D+1}$ are
pictured in Fig.7; the vertex for $\bar \phi^{D+1}$ can be easily found by conjugation).

\begin{figure}
\centering
\centering
\includegraphics[angle=0, width=14cm, height=6cm]{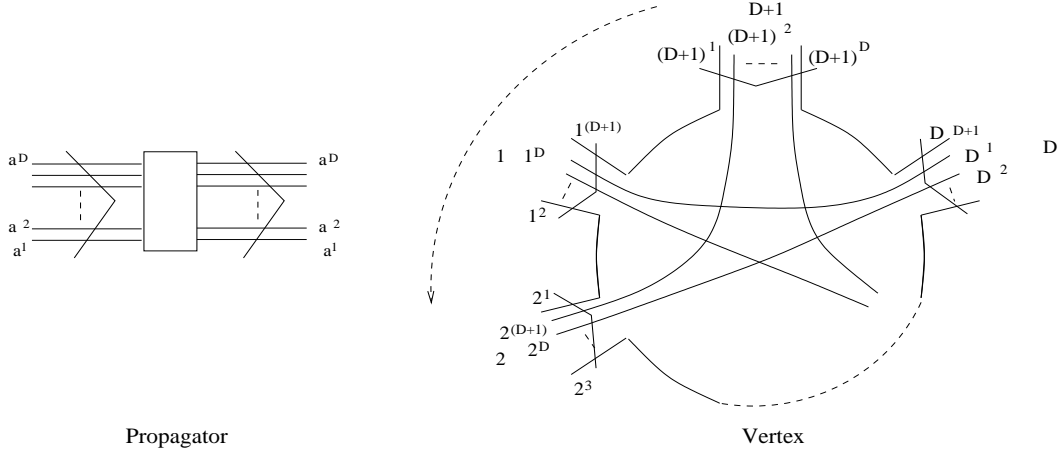}
\caption{
{\small Propagator and vertex $\phi^{D+1}$ in $D$ dimensional GFT. }}
\end{figure}
\noindent{\bf Perturbative bounds.}
An extension of Theorem \ref{theo1}  can be easily realized here in any dimension
using similar conditions as in Lemma \ref{lem1} for any $N$-point function
(the argument on the parity of the number of lines in complete contraction procedure
will still hold here). Since there is no generalized tadpole again in the
$D$ dimensional theory, we are able
to still provide a exhaustive sequence of cuts for any colored graph. From
this point, we can formulate the following statement:
\begin{theorem}
There exists a constant $K$ such that for any connected colored vacuum graph $\mathcal{G}$ of the
D dimensional GFT model with $n$ internal vertices, we have
\begin{equation}
 |\mathcal{A}_\mathcal{G}|\leq K^n \Lambda^{3(D-1)(D-2)n/4 + 3(D-1)}.
\end{equation}
This bound is optimal in the sense that there exists a graph $\mathcal{G}_n$ with $n$ internal vertices
 such that  $|\mathcal{A}_\mathcal{G}| \simeq  K^n \Lambda^{3(D-1)(D-2)n/4 + 3(D-1)}$.
\end{theorem}
{\bf Proof.} Given $0<p\leq D$, the generalized vertex operator is an operator with $D+1-p$ legs in
a part $A$ and remaining $p$ legs in a part $B$, for a $(A,B)$-cut for $\mathcal{G}$ (see Fig.8).
\begin{figure}
\centering
   \begin{minipage}[t]{0.6\textwidth}
\centering
\includegraphics[angle=0, width=8cm, height=4cm]{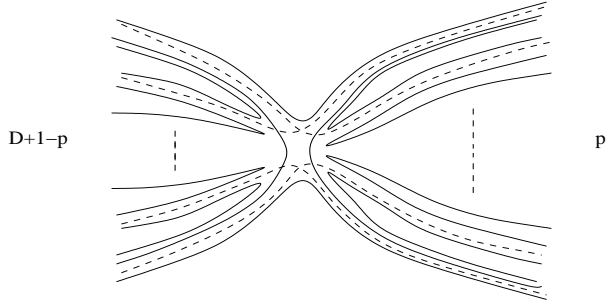}
\caption{
{\small The operator $O(D+1-p,p)$. }}
\end{minipage}
\end{figure}
From the generalized formula for
\begin{eqnarray}
\text{tr} (O_{D+1-p, p} O_{p, D+1-p})^{2n}= \Lambda^{
3 \left[\left( (D-p-1)(D-p)+ (p-1)(p-2)\right)n + (D+1-p)p -1\right]
},
\end{eqnarray}
we can determine the bound on the norm of the operator $O_{ D+1-p,p}$
as
\begin{eqnarray}
|| O_{ D+1-p,p} || \leq \Lambda^{\frac{3(D-1)(D-2)- 2(p-1)(D-p)}{4}}.
\end{eqnarray}
Then the maximum of the bound occurs for $p=1$, giving an operator $O_{D,1}$.
For this operator  we have
\begin{equation}
|| O_{D,1} || \leq (\Lambda^{3/2})^{\frac{(D-1)(D-2)}{2}} . \qquad \square
\label{od1}
\end{equation}

Setting $D=4$ and $p=1$, (\ref{od1}) recovers the bounds of $O_{14} $ (\ref{o14}).
In the specific $D=3$ dimensional case, 
(\ref{od1}) reduces to $\Lambda^{3/2}$ bound of the vertex operators of  Boulatov's model 
as established in \cite{MNRS}.

\section{Integration of two fields: first steps of constructive analysis}
\label{sect3}

A main interesting feature of the colored theory, is the possibility of an explicit integration of
two fields leading to a determinant as in the Matthews-Salam formalism
\cite{msalam}. We explore this possibility in four
dimensions in this subsection, but the result again is general. 
 Let us start by writing the vertex terms in the form (setting henceforth
$\lambda_1 = \lambda_2= \lambda$)
\begin{eqnarray}
 S_{V} &=& \lambda \left\{  \int \prod_{i=1}^{4} dg_i dg'_i \;
\phi^1_{1,2,3,4}\left[ \delta(g_1(g'_1)^{-1}) \;
\int dg_5dg_6dg_7 \; \phi^5_{4,2',5,6}\, \phi^4_{6,3,3',7}\, \phi^3_{7,5,2,4'}  \right] \phi^2_{4',3',2',1'} \right. \cr
& +& \left. \int\prod_{i=1}^{4}dg_i dg'_i \;
{\bar\phi}^1_{4',3',2',1'} \left[ \delta(g'_4(g_4)^{-1})
\int dg_5dg_6dg_7 \, {\bar\phi}^5_{1',3,5,6}\, {\bar\phi}^4_{6,2,2',7}
\, {\bar\phi}^3_{7,5,3',1} \right] {\bar\phi}^2_{1,2,3,4}\right\},
\end{eqnarray}
such that the following operators
\begin{eqnarray}
H(g_1,g_2,g_3,g_4; g'_4,g'_3,g'_2,g'_1) &=&
 \delta(g_1(g'_1)^{-1})
\int dg_5dg_6dg_7 \; \phi^5_{4,2',5,6}\, \phi^4_{6,3,3',7}\, \phi^3_{7,5,2,4'}  \;, \\
H^* (g_1,g_2,g_3,g_4; g'_4,g'_3,g'_2,g'_1) &=&
\delta(g_4(g'_4)^{-1})  \int dg_5dg_6dg_7 \;{\bar\phi}^5_{1',3,5,6}\, {\bar\phi}^4_{6,2,2',7}
\, {\bar\phi}^3_{7,5,3',1}
\end{eqnarray}
allow us to express the partition function (\ref{parti}) as
\begin{eqnarray}
Z(\lambda)&=& \int  d \mu_{C}[\bar\phi,\phi]\;
\exp[ - \lambda [ \; \int \prod_{i=1}^{4} dg_i dg'_i \;\;
\phi^1_{1,2,3,4}\; H(g_1,g_2,g_3,g_4;g'_4,g'_3,g'_2,g'_1)\;
\phi^2_{4',3',2',1'} \crcr
&&
+ {\bar\phi}^2_{1,2,3,4}\; H^* (g_1,g_2,g_3,g_4;g'_4,g'_3,g'_2,g'_1) \;
{\bar\phi}^1_{4',3',2',1'}]].
\end{eqnarray}
Note the important property of $H^*$ to be the adjoint of $H$. Indeed, a quick inspection
shows that $H^* = (\bar H)^t$ which can be checked by a complex conjugation
of the fields and a symmetry of arguments such that $1 \to 4$ and $2 \to 3$.

The integration of this function follows standard techniques in quantum field theory.
We can introduce the vector $v= ({\rm Re}\phi^1,{\rm Im}\phi^1, {\rm Re}\phi^2,{\rm Im}\phi^2 )$
and its transpose $v^t$ and such that $\phi^1 H  \phi^2$ and $\bar\phi^1 \bar H \bar \phi^2$
and the mass terms $\bar\phi^{\ell=1,2}_{1,2,3,4}\phi^{\ell=1,2}_{4,3,2,1}$ can
be expressed as in the matrix form
\begin{equation}
Z(\lambda) =  \int  d \mu_{C}[\bar\phi,\phi]\;
e^{- \lambda v^t\, A \,  v  },
\end{equation}
where the matrix operator $A$ can be expressed as
\begin{equation}
A = \left(\begin{array}{cccc}
0&0& H&i H\\
0&0&i H&-H\\
H^*&-i H^*&0&0\\
-i H^*&-H^*&0&0
\end{array}\right).
\end{equation}
After integration over the colors 1 and 2, using the normalized Gaussian  measure
$$d\mu_C[\bar\phi, \phi]=  d\mu'_{C'} [\bar\phi^{1,2};\phi^{1,2}] d\mu''_{C''} [\bar\phi^{3,4,5} , \phi^{3,4,5}] ,$$
we get
\begin{equation}
Z(\lambda) = \int d\mu''_{C''} [\bar\phi^{3,4,5} , \phi^{3,4,5}]  \; K [ \det (1 + \lambda C A)] ^{-1}=
 \int d\mu''_{C''} [\bar\phi^{3,4,5} , \phi^{3,4,5}]  \;  K e^{-\text{tr} \log(1+ \lambda CA)},
\label{det}
\end{equation}
where $K$ is an unessential normalization constant that we omit in the sequel.
Mainly the operator product $ CA$
can be calculated by composing $C H$ and $C H^*$. We obtain for $C H$ and $C H^*$, respectively,
the following operators defined by their kernel
\begin{eqnarray}
&&{\mathcal H}(g_1,g_2,g_3,g_4;g'_4,g'_3,g'_2,g'_1)= \cr
&&
 \int \prod_{i=1}^{4} dg''_i\;
C(g_1,g_2,g_3,g_4;g''_4,g''_3,g''_2,g''_1) H(g''_1,g''_2,g''_3,g''_4; g'_4,g'_3,g'_2,g'_1)\\
&=&
\int dg_5dg_6dg_7 \int dh \,\delta(g_1 h (g'_1)^{-1}) \; \phi^5(g_4h,g'_2,g_5,g_6)\; \phi^4(g_6,g_3h,g'_3,g_7)\;
\phi^3(g_7,g_5,g_2h,g'_4) \;, \cr
&&\crcr
&&
{\mathcal H}^*(g_1,g_2,g_3,g_4;g'_4,g'_3,g'_2,g'_1) =\cr
&& \int \prod_{i=1}^{4} dg''_i\;\;
C(g_1,g_2,g_3,g_4;g''_4,g''_3,g''_2,g''_1) H^*(g''_1,g''_2,g''_3,g''_4; g'_4,g'_3,g'_2,g'_1)\\
&=&
\int dg_5dg_6dg_7  \int dh \,\delta(g_4 h (g'_4)^{-1})\; \phi^5(g_1',g_3h,g_5,g_6)\; \phi^4(g_6,g_2',g_2h,g_7)\;
\phi^3(g_7,g_5,g_3',g_1h).
\nonumber
\end{eqnarray}
It is convenient to see $C A$ as a sum of two matrices  $\mathbb{H} + \mathbb{H}^*$
which are defined by the off block diagonal matrices of $C A$ involving at each matrix
element  $\mathcal{H}$ or ${\mathcal H}^*$.
Then the determinant integrant of (\ref{det}) can be expressed, up to a constant, more simply by
\begin{eqnarray}
&& e^{-\text{tr}\log (1+ \lambda (\mathbb{H} + \mathbb{H}^*)) } =
e^{+\text{tr} \sum_{n=1}^\infty \frac{(-\lambda)^{n}}{n} (\mathbb{H} + \mathbb{H}^*)^n}
\cr
 && =e^{+\text{tr} \sum_{p=1}^\infty \frac{\lambda^{2p}}{2p}  ([\mathbb{H} + \mathbb{H}^*)^2])^p}
=e^{\frac{1}{2}\text{tr} \sum_{p=1}^\infty \frac{\lambda^{2p}}{p} (Q)^p}=
e^{-\frac{1}{2}\text{tr} \log(1-\lambda^2 Q)}
\end{eqnarray}
where  we have used the fact that $\text{tr} ((\mathbb{H} + \mathbb{H}^*)^{2p+1})=0$ for all $p$,
and $Q:= (\mathbb{H} + \mathbb{H}^*)^2 = 
\mathbb{H} \mathbb{H}^* +  \mathbb{H}^*  \mathbb{H}$,
since  $\mathbb{H}^2 = 0 = (\mathbb{H}^*)^{2}$. The operator $Q$ is Hermitian and is given as
\begin{eqnarray}
 Q &=&
\left(\begin{array}{cccc}
2{\mathcal H} {\mathcal H}^*&-2i  {\mathcal H}{\mathcal H}^*&0&0\\
2i  {\mathcal H}  {\mathcal H}^*&2  {\mathcal H}{\mathcal H}^*&0&0\\
0&0& 2{\mathcal H}^* {\mathcal H}& 2i {\mathcal H}^*  {\mathcal H}\\
0&0&-2i  {\mathcal H}^* {\mathcal H}& 2 {\mathcal H}^* {\mathcal H}
\end{array}\right).
\end{eqnarray}
Thus, $Q$ has {\emph {positive}} real eigenvalues and $-\lambda^2 Q$ is positive for $\lambda = ic$, $c\in \mathbb{R}$.

The next purpose is to investigate a bound on the radius of the series
$F(\lambda)= \log Z(\lambda)$ denoting the free energy and computing 
the amplitude sum of connected Feynman graphs. 
For this, we will use a cactus expansion \cite{riv2}  and  
the Brydges-Kennedy forest formula (see \cite{riv3} and references therein; for a short pedagogical
approach see \cite{MNRS}) on the partition function $Z(\lambda)$. 
Expanding $e^{-\frac{1}{2}\text{tr} \log(1-\lambda^2 Q)}$
in terms of $V_\lambda= -\frac{1}{2}\text{tr} \log(1-\lambda^2 Q)$, using the replica trick, and then applying the
Brydges-Kennedy formula, one comes to 
\begin{eqnarray}
\label{lvertex}
Z(\lambda)&=& \int d\nu_{C} [\bar\phi^{5,4,3},\phi^{5,4,3}]\;e^{-\frac{1}{2}\text{tr} \log(1-\lambda^2 Q)}\cr
&=&
\sum_{n=0}^{\infty} \frac{1}{n!}\sum_{F\in\mathcal{F}_n}\left(\prod_{l\in F}\int_0^1dh_l\right)\left(\prod_{l\in F}
\frac{\partial}{\partial h_l}\right)\int d\nu^{\boldsymbol{h^F}}_n(\sigma_1,\dots,\sigma_n)\prod_{v=1}^nV_{\lambda}(\sigma_v),
\end{eqnarray}
where we have changed the notation $d\nu_{C} =d\mu''_{C''}$ and each of the $\sigma_i$ represent
an independent copy of the six fields $(\bar\phi^{5,4,3},\phi^{5,4,3})_i$;
the second sum is over the set $\mathcal{F}_n$ of forests  built on $n$ points or vertices;
the product is over lines $l$ in a given forest $F$; 
 $\boldsymbol{h^F}$ is a $n(n-1)/2$-tuple with element $h_l^F= \min_{p} h_p$, 
where $p$ take values in the unique path in $F$ connecting the source $s(l)$ 
and target $t(l)$ of a line $l\in F$, if such a path exists, otherwise $h_l^F=0$.
It is well known that the summand factorizes along connected components of each forest.
Therefore $\log Z(\lambda)$ is given by the same series in terms of trees (connected forests) as
\begin{eqnarray}
\label{toytrees}
F(\lambda)=\sum_{n=1}^{\infty}\frac{1}{n!}\sum_{T\in \mathcal{T}_n}\left(\prod_{l\in T}\int_0^1dh_l\right)\left(\prod_{l\in T}
\frac{\partial}{\partial h_l}\right)\int d\nu^{\boldsymbol{h^T}}_n(\sigma_1,\dots,\sigma_n)\prod_{v=1}^nV_{\lambda}(\sigma_v).
\end{eqnarray}
The trees $T$ join the new vertices $V_{\lambda}(\sigma_{v})$'s often called ``loop vertices''. 
The covariance of $ d\nu^{\boldsymbol{h^T}}_n(\sigma_1,\dots,\sigma_n)$ can be expressed by,
\begin{eqnarray}
 C^{\boldsymbol{h^T}}_{ij;\, ab}(g^i;g^j) = \left\{
                  \begin{array}{cc}
                  1     &{\rm if}\;\; i=j \;\; {\rm and} \;\;a=b\\
                  \delta_{ab}\; h^T_l \; C(g^{i};g^{j}) &{\rm if}\;\; i\neq j\\
                     0& {\rm otherwise}
                           \end{array}
 \right. 
\end{eqnarray}
where $l=\{ij\}$ denotes the line with source $s(l)=i$ and target $t(l)=j$.
We use also the notation $(g^i,g^j)=((g^i)_{k=1,2,3,4},(g^j)_{k=1,2,3,4})\in G^{4\times 4}$, 
$a$ and $b$ are color indices such that formally, we have
\begin{eqnarray}
 d\nu^{\boldsymbol{h^T}}_n(\sigma_1,\dots,\sigma_n)=
e^{\; \int dg dg'\; \sum_{i,j=1}^n\sum_{a,b=1}^5 
\frac{\delta}{\delta\bar\phi^{a}_{i}(g^i)}
C^{\boldsymbol{h^T}}_{ij; ab}(g^i,g^j) 
\frac{\delta}{\delta\phi^{b}_{j}(g^j)}}.
\end{eqnarray}
The partial derivative $\partial /\partial h_l$ in (\ref{toytrees}) acts on the measure
 $d\nu^{\boldsymbol{h^T}}_n(\sigma_1,\dots,\sigma_n)$ and one obtains
\begin{eqnarray}\label{cactusBFL}
F(\lambda) &=&\sum_{n=1}^{\infty}\frac{1}{n!}\sum_{T\in\mathcal{T}_n}\left(\prod_{l\in T}\int_0^1dh_l\right)\int d\nu^{\boldsymbol{h^T}}_n(\sigma_1,\dots,\sigma_n)\\
&&\prod_{l\in T}\int d^4g^{s(l)}d^4g^{t(l)}\   C(g^{s(l)};g^{t(l)})\frac{\delta^2}{\delta\bar\phi^{a(l)}_{s(l)}(g^{s(l)})\delta\phi^{a(l)}_{t(l)}(g^{t(l)})}
\prod_{v=1}^nV_{\lambda}(\sigma_v),
\nonumber
\end{eqnarray}
with the color  index $a(l)$ denoting the color of the line $l$. 

Let us denote by $k_v$ the coordination number of a given loop vertex 
$V_{\lambda}(\sigma_v)$ hooked to the tree $T$ by half lines $l_v=1, \ldots, k_v$ such that
$s(l_v)=v$ or $t(l_v)=v$. Since each half line corresponds to a derivative,
we can decompose the  $k_v$ derivatives in two parts $p_v+q_v= k_v$ such that,
up to inessential constants
\begin{eqnarray}
&&{\mathcal T}(\lambda;k_v)= \prod_{l_v^1=1}^{p_v}
\frac{\delta}{\delta\bar\phi^{a(l^1_v)}_{v}(g^{l^1_v})}
 \prod_{l_v^2=1}^{q_v}\frac{\delta}{\delta\phi^{a(l^2_v)}_{v}(g^{l^2_v})}
\left(-\frac{1}{2} {\rm tr} \log (1-\lambda^2 Q[\bar\phi^{5,4,3}_v;\phi^{5,4,3}_v])\right) 
\label{ultr}\\
&& =  {\rm tr}
\sum_{\stackrel{r}{\stackrel{p_r + q_r>0}{\sum p_r =p_v; \sum q_r=q_v}}}
\prod_{j=1}^r\left\{ \frac{\delta^{p_j+q_j}\; \lambda^2 Q}{\prod_{l^{(1;j)}_v=1}^{p_j} \delta\bar\phi^{a(l^{(1;j)}_v)}_{v}(g^{l^{(1;j)}_v})\,
\prod_{l^{(2;j)}_v=1}^{q_j} \delta\phi^{a(l^{(2;j)}_v)}_{v}(g^{l^{(2;j)}_v})} \frac{1}{1- \lambda^2 Q}
\right\}  .
\nonumber
\end{eqnarray}
Note that, as previously mentionned the coupling constant $\lambda = ic$ is a pure imaginary complex number
so that the denominator $1+ c^2 Q$ is positive. 

A rigorous bound for (\ref{ultr}) is complicated to determine. However an
encouraging remark is to consider the constant field modes or constant fields
themselves (even if the physical relevance  of these ``background'' modes is not clear at this stage). 
If we restrict to these constant modes, then (\ref{ultr}) can be bounded as follows.
Noting that $Q$ is a polynomial in the fields $\phi^a$ of degree six,
the product of derivatives acting on it behaves like
\begin{eqnarray}
{\mathcal T}(\lambda;k_v) \leq  \sum_{\stackrel{r}{\stackrel{p_r + q_r>0}{\sum p_r =p_v; \sum q_r=q_v}}}
\prod_{j=1}^r 
\frac{|\lambda^{1/3}\phi|^{6-p_j-q_j} (\lambda^{1/3})^{p_j+q_j}}{1+ |\lambda^{1/3}\phi|^6}
\leq  \sum_{\stackrel{r}{\stackrel{p_r + q_r>0}{\sum p_r =p_v; \sum q_r=q_v}}}
\prod_{j=1}^r  (\lambda^{1/3})^{p_j+q_j} .
\end{eqnarray}

Note that similar formulas exist in dimension $D$ with $Q$ a polynomial of degree $2(D-1)$.
Hence this method opens a constructive perspective for colored GFT in any dimension.
This perspective is distinct or could be a complement to the Freidel-Louapre approach.

\section{Conclusion}

Perturbative bounds of a general vacuum graph of colored GFT have been obtained in any dimension. 
We have shown that, in dimension $D$,  the scaling of the amplitude of any vacuum graph,  in the ``ultraspin'' cutoff
$\Lambda$, behaves like $K^n \Lambda^{3(D-1)(D-2)n/4+ 3(D-1)}$ where $n$ is the number of vertices.
In addition, it has been revealed that, by an integration of two colors, the model is positive if
the coupling constant is purely imaginary. We did not need further regularization procedure,
such as an inclusion of a  Freidel-Louapre interaction term which in return may be 
more divergent than the ordinary theory. Using a loop vertex expansion, 
we have reached the property that for at least the constant modes of the fields
the forest-tree formula leads to a convergent series (which should be
the Borel-Le Roy sum of appropriate order of the ordinary perturbation series).
These results are encouraging for a constructive program. 

The bounds obtained in this paper for colored GFT models 
should now be completed into a more precise power counting and scaling analysis.
They should also be extended to the physically more interesting 
models in particular the so called EPRL-FK model  \cite{engle,fkras}.
We recall that
the most simple divergencies of the EPRL-FK model have been studied in \cite{prs}. 
This work analyzes
the elementary ``bubble'' (built out of two vertices) corresponding to the one-loop
self-energy correction and the ``ball'' (built out of five vertices), corresponding
to the one-loop vertex correction.  The authors prove that putting the external legs at zero
spin, the degree of divergence in the cutoff of the ball could be logarithmic.
This result sounds very promising from the renormalization point of view.

The first step in this program is to establish the power counting
of a simplified colored Ooguri model with commutative group, which we
call ``linearized'' colored Ooguri model. We checked that 
power counting in this case is given by a homology formula
\cite{gurau,bmr}.  Then we intend to perform a similar 
analysis of the colored linearized EPRL-FK  model, and find out the analogs
of the multiscale renormalization group analysis in this context.
This should help for the more complicated study of the ``non-linear`` models
in which homotopy rather than homology should ultimately govern the power counting.

\section*{Acknowledgments}

The work of J.B.G. is supported by the Laboratoire de Physique Th\'eorique d'Orsay
(LPTO, Universit\'e Paris Sud XI) and the Association pour la Promotion Scientifique 
de l'Afrique (APSA). Discussions with R. Gurau are gratefully acknowledged, which 
in particular lead to the representation of Section 3.

\section{Appendix: The case of tadpoles}
\begin{figure}
\begin{center}
\includegraphics[angle=0, width=10cm, height=3.5cm]{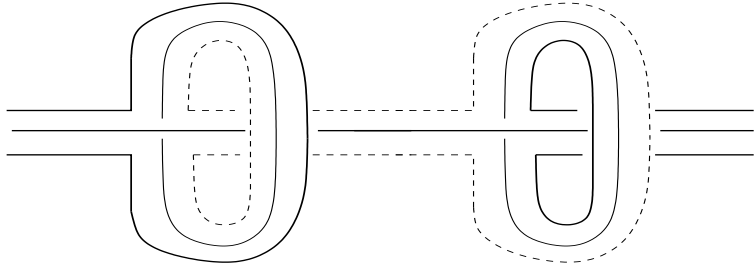}
\caption{
{\small Nonplanar tadpoles with their additional divergence (dashed line) when inserted in
more involved graphs. }}
\end{center}
\end{figure}

In this appendix, we give some precisions on the scaling properties
of  amplitudes of graphs containing generalized tadpoles as occur in the non colored GFT.
We will restrict the discussion to the case of dimension 3 \cite{MNRS,fgo} but 
similar properties definitely appear in greater dimension. 

Referring to \cite{MNRS}, the vertex operators of Boulatov's model 
are bounded by $\Lambda^{3/2}$
such that any connected vacuum graph without generalized tadpoles is bounded by
$K^n\Lambda^{3n/2}$ (omitting the overall trace factor), $n$ being as usual the number of vertices. 
However, nothing more can be said in the case of graphs with generalized tadpoles.
For instance, there exist non-planar tadpoles (see Fig.9) which contribute
more than $\Lambda^{3/2}$ per vertex. 
Indeed, in the typical situation of Fig.9, these tadpoles cost a factor $\Lambda^{3/2 + 3/2}$
per vertex, and violate the ordinary vertex bound. 

Finally, one concludes that the bounds of \cite{MNRS} are correct for vacuum graphs without generalized tadpoles,
such as the vacuum graphs that occur in the colored models,
but that, when tadpoles are present, they are wrong for the most general model.
In any non-colored case and any dimension, we can expect similar features.


\begin{thebibliography}{99}

\bibitem{boul}
D. V. Boulatov, Mod. Phys. Lett. {\bf  A7} (1992) 1629;
eprint {\tt hep-th/9202074}.


\bibitem{oogur}
H. Ooguri, Mod. Phys. Lett. {\bf  A7} (1992) 2799;
eprint {\tt hep-th/9205090}.

\bibitem{Freidel}
L. Freidel, Int. J. Theor. Phys. {\bf 44} (2005) 1769;
eprint {\tt hep-th/0505016}.

\bibitem{oriti}
D. Oriti, ``The group field theory approach to quantum gravity,''
 eprint {\tt gr-qc/0607032} (2006).

\bibitem{rovel}
C. Rovelli, {\it Quantum Gravity}
(Cambridge University Press, Cambridge, 2004).

\bibitem{thiem}
T. Thiemann, {\it Modern canonical quantum general relativity}
(Cambridge University Press, Cambridge 2007).

\bibitem{fgo}
L. Freidel, R. Gurau and D. Oriti,
Phys. Rev. {\bf D80} (2009) 044007;
eprint {\tt 0905.3772[hep-th]}.

\bibitem{MNRS}
J. Magnen, K. Noui, V. Rivasseau and M. Smerlak,
``Scaling behaviour of three dimensional group field theory,''
eprint {\tt 0906.5477[hep-th]} (2009).

\bibitem{riv}
V. Rivasseau, {\it Noncommutative renormalization}, Poincar\'e Seminar X,
 ``Espaces Quantiques'', ed. B. Duplantier {\it et al}, (2007) 15-95;  eprint {\tt 0705.0705[hep-th]}.

\bibitem{prs}
C. Perini, C. Rovelli and S. Speziale,
``Self-energy and vertex radiative corrections in LQG,''
eprint {\tt  0810.1714[gr-qc]} (2008).

\bibitem{barret}
J. W. Barrett,  R. J. Dowdall,  W. J. Fairbairn,  H. Gomes and
F. Hellmann, ``Asymptotic analysis of the EPRL four-simplex amplitude,''
eprint {\tt 0902.1170[gr-qc]} (2009).


\bibitem{flouapr}
L. Freidel and  D. Louapre,
Phys. Rev. {\bf D68} (2003) 104004;
 eprint {\tt hep-th/0211026}.

\bibitem{gurau}
R. Gurau, ``Colored group field theory,''
eprint {\tt 0907.2582[hep-th]} (2009).

\bibitem{msalam}
P. T. Matthews and A. Salam, 
Nuovo Cimento {\bf 12} (1954)  563;
{\it ibid} {\bf 2} (1955) 120.


\bibitem{fkras}
L. Freidel and K. Krasnov,
Class. Quant. Grav.  {\bf 25} (2008) 125018;
eprint {\tt 0708.1595[gr-qc]}.

\bibitem{engle}
J. Engle, E. Livine, R. Pereira and C. Rovelli,
Nucl. Phys. {\bf B799} (2008) 136;
eprint {\tt  0711.0146[gr-qc]}.

\bibitem{riv2}
V. Rivasseau, JHEP {\bf 9} (2007) 008; eprint {\tt 0706.1224[hep-th]}.

\bibitem{riv3} 
V. Rivasseau, {\it From Perturbative to Constructive Renormalization} (Princeton University Press, Princeton, 1991);
A. Abdesselam and V. Rivasseau, {\it Trees, forests and jungles: a botanical
garden for cluster expansions}, in Constructive Physics, ed. V. Rivasseau,
Lecture Notes in Physics 446, Springer Verlag, 1995.


\bibitem{bmr} J. Ben Geloun, J. Magnen and V. Rivasseau,
{\it in progress}.

\end{thebibliography}
\end{document}